\documentclass[12pt]{iopart}
\usepackage{geometry}                
\usepackage{graphicx}
\usepackage{amssymb}
\usepackage{epstopdf}
\begin{document}
\title{Particle sizing by dynamic light scattering: non-linear cumulant analysis}
\author{Alastair G. Mailer, Paul S. Clegg, and Peter N. Pusey}
\address{SUPA, School of Physics and Astronomy, University of Edinburgh,
Peter Guthrie Tait Road, Edinburgh, EH9 3FD} 
\ead{a.mailer@physics.org}

\begin{abstract}
We revisit the method of cumulants for analysing dynamic light scattering data in particle sizing applications.  Here the data, in the form of the time correlation function of scattered light, is written as a series involving the first few cumulants (or moments) of the distribution of particle diffusion constants.  Frisken (2001 Applied Optics \textbf{40} 4087, \cite{frisken}) has pointed out that, despite greater computational complexity, a non-linear, iterative, analysis of the data has advantages over the linear least-squares analysis used originally.  In order to explore further the potential and limitations of cumulant methods we analyse, by both linear and non-linear methods, computer-generated data with realistic `noise', where the parameters of the distribution can be set explicitly.  We find that, with modern computers, non-linear analysis is straightforward and robust.  The mean and variance of the distribution of diffusion constants can be obtained quite accurately for distributions of width (standard deviation/mean) up to about 0.6, but there appears to be little prospect of obtaining meaningful higher moments.
\end{abstract}

\maketitle

\section{Introduction}

Small particles of many kinds occur naturally in biological systems and the environment, and are used at some stage in many industrial processes~\cite{Merkus}.  Frequently the size of the particles is crucial to their function.  Thus measuring particle size is important, and many techniques have been developed for this purpose~\cite{Merkus}.  One such technique, dynamic light scattering (DLS), can be applied to nanometric particles, size from a few nm to about 1\,$\mu$m, which can be suspended in a liquid. DLS has the advantages of being quick and reproducible and of providing well-defined (though often limited) information about the particles.

In DLS, coherent laser light scattered by a particle suspension forms a random diffraction pattern that fluctuates in time as the particles move in Brownian diffusion~\cite{berne, pusey1, iso}.  For equal-sized particles, analysis of the time dependence of the scattered light yields the particles' diffusion constant from which its size can be calculated.  In the more common situation where the suspension is polydisperse --- there is a distribution of particle sizes --- DLS yields the Laplace transform of the distribution of diffusion coefficients.  Thus, in principle, the latter, usually the quantity of interest, can be obtained by inverse Laplace transformation of the data.  A variety of techniques have been used to perform this operation, including exponential sampling \cite{ostrowsky}, regularization \cite{provencher}, maximum entropy \cite{livesey}, maximum likelihood \cite{sun} and non-negatively constrained least squares \cite{zhu}.  Because inverse Laplace transformation is particularly sensitive to inevitable experimental uncertainties in the data, these techniques are best suited to broad distributions (which may be multi-modal). Furthermore, the techniques can be quite complicated to operate and may require the input of prior information about the sample such as minimum and maximum particle size.  A survey and critique of these methods, 20 years old but still valuable, was given by Finsy~\cite{Finsy}.

A simpler approach to DLS data analysis, which was in fact the first one to go beyond fitting a single exponential, is the so-called method of cumulants~\cite{koppel, pusey2, brown}.  This provides a few lower moments, or cumulants, of the distribution of diffusion coefficients, and is the topic addressed in this paper. In the early days of DLS, when computing power was limited, cumulant analysis used a linear fitting method which did not require an iterative program.  Later Frisken \cite{frisken} pointed out that a much more versatile non-linear, iterative, fitting procedure is possible with modern computers.  Frisken \cite{frisken, patty} and others \cite{hassan} demonstrated the value of this approach to analyze real experimental systems.  Here we use realistic computer-generated data, where the parameters of the size distribution can be set explicitly, to evaluate further the potential and limitations of non-linear cumulant analysis.

In the next section we describe the background to the current situation.  While much of this material has appeared before, we believe that it is helpful to provide a coherent account here.  Section \ref{methods-sec} describes generation of the synthetic data and the analysis methods used.  In section \ref{results-sec} we present the results which are discussed in section \ref{discuss-sec}.  We find that, if the distribution of diffusion coefficients is not too broad, non-linear cumulant analysis offers a straightforward and robust method for determining its mean and variance.  However, the prospects for obtaining higher moments are not promising. 

\section{Background}

General references for this section are \cite{berne, pusey1, koppel, pusey2, brown}.  DLS measures the normalised time correlation function $g^{(2)}(\tau)$ of the scattered light intensity $I$:
\begin{equation}
g^{(2)}(\tau) = \frac{ \langle I(0) I(\tau) \rangle }{ \langle I \rangle^2 },
\label{first-dls-eqn}
\end{equation}
where $\tau$ is the correlation delay time.  In many cases the intensity correlation function can be written in terms of the correlation function $g^{(1)}(\tau)$ of the scattered light field through the so-called Siegert relation \cite{siegert,pusey1}:
\begin{equation}
g^{(2)}(\tau) = 1 + \beta \left[ g^{(1)}(\tau) \right]^2,
\label{siegert-eqn}
\end{equation}
where $\beta$ is the coherence factor, determined largely by the ratio of the detector area to the coherence area of the scattered light; $\beta$ is usually regarded as an unknown parameter to be fitted in the data analysis.  In the simplest case of a dilute suspension of identical spherical particles in Brownian motion, $ g^{(1)}(\tau)$ is given by
\begin{equation}
g^{(1)}(\tau) = \exp (-\Gamma \tau )
\label{mono-g1-eqn}
\end{equation}
where the decay rate $\Gamma$ is
\begin{equation}
\Gamma = Dq^2,
\label{diff-eqn}
\end{equation}
$D$ is the translational diffusion coefficient of the particles and $q$ is the scattering vector (set by the scattering angle $\theta$ and the wavelength $\lambda$ of the light in the sample through $q = (4 \pi / \lambda) \sin (\theta / 2)$).  In turn, for spherical particles, $D$ is given by the Stokes-Einstein relation
\begin{equation}
D = \frac{k_B T}{6 \pi \eta R}
\label{last-dls-eqn}
\end{equation}
where $k_B T $ is the thermal energy, $\eta$ the viscosity of the solvent and $R$ the particles' radius.  Equations (\ref{first-dls-eqn}--\ref{last-dls-eqn}) form the basis of particle sizing by DLS.

When a sample is polydisperse, containing particles of different sizes, each species gives rise to its own exponential decay in the field correlation function so that
\begin{equation}
g^{(1)}(\tau) = \int G(\Gamma)\exp(-\Gamma \tau) d\Gamma,
\label{poly-g1-eqn}
\end{equation}
where $G(\Gamma)$ is the normalized distribution of decay rates $\Gamma$ ($\int G(\Gamma) d\Gamma = 1$).  Thus $g^{(1)}(\tau)$, obtained from the measurement of $g^{(2)}(\tau) $ through Eq. (\ref{siegert-eqn}), is the Laplace transform of $G(\Gamma)$.  In principle, therefore, the latter can be obtained by inverse Laplace transformation of the data. In practice, inverse Laplace transformation is an ill-conditioned problem in the sense that it converts small uncertainties in the data into large uncertainties in the recovered $G(\Gamma)$ \cite{mcwhirter}.  Put another way, unless there is a wide spread of particle size, the sum of exponentials implied by Eq. (\ref{poly-g1-eqn}) looks not too different from a single, average, exponential.

This limitation can be recognized and exploited by writing
\begin{equation}
\exp (-\Gamma \tau ) = \exp (-\bar{\Gamma} \tau) \exp \left[ -(\Gamma - \bar{\Gamma}) \tau \right],
\end{equation}
where $\bar{\Gamma}$ is the mean value of $G(\Gamma)$,
\begin{equation}
\bar{\Gamma} = \int \Gamma G(\Gamma) d\Gamma,
\end{equation}
and expanding the second exponential to give, in Eq. (\ref{poly-g1-eqn}),
\begin{equation}
g^{(1)}(\tau) = \exp (-\bar{\Gamma} \tau) \left[ 1 + \frac{1}{2} \mu_2 \tau^2 - \frac{1}{3!} \mu_3\tau^3 + \frac{1}{4!} \mu_4 \tau^4 - \dots \right],
\label{g1-moments-eqn}
\end{equation}
where
\begin{equation}
\mu_n = \int (\Gamma - \bar{\Gamma})^n G(\Gamma) d\Gamma
\label{mun-eqn}
\end{equation}
are the central moments of the distribution of decay rates (the moments about the mean).  Equation (\ref{g1-moments-eqn}) shows clearly how DLS data can be represented by an average exponential with correction terms that depend on $G(\Gamma)$ (and hence on the particle size distribution).  For a reasonably narrow distribution of decay rates and a reasonable range of scaled delay time $\bar{\Gamma}\tau$, the higher-order terms in Eq. (\ref{g1-moments-eqn}) become increasingly unimportant.  Rewriting Eq. (\ref{g1-moments-eqn}) in terms of scaled time,
\begin{equation}
\fl g^{(1)}(\tau) = \exp (-\bar{\Gamma} \tau) \left[ 
1 
+ \frac{1}{2} \frac{\mu_2}{\bar{\Gamma}^2} (\bar{\Gamma} \tau)^2
- \frac{1}{3!} \frac{\mu_3}{\bar{\Gamma}^3} (\bar{\Gamma} \tau)^3
+ \frac{1}{4!} \frac{\mu_4}{\bar{\Gamma}^4} (\bar{\Gamma} \tau)^4
- \dots
\right],
\end{equation}
shows that $\mu_2 / \bar{\Gamma}^2 $, the normalized variance of $G(\Gamma)$, is the simplest measure of the departure of $g^{(1)} (\tau) $ from a single exponential.   With Eq. (\ref{siegert-eqn}), Eq. (\ref{g1-moments-eqn}) can also be written as
\begin{equation}
\fl \ln \sqrt{g^{(2)}(\tau) - 1 } = \frac{1}{2} \ln \beta 
- \bar{\Gamma} \tau 
+ \frac{1}{2} \mu_2 \tau^2
- \frac{1}{3!} \mu_3 \tau^3 
+ \frac{1}{4!} (\mu_4 - 3\mu_2^2) \tau^4
-\dots,
\label{linear-g1-fit-eqn}
\end{equation}	
showing further how non-exponentiality appears as departure from linearity in a semi-logarithmic plot of the data (see Fig. \ref{sim-res-fig}).

The original method of cumulants \cite{koppel, pusey2, brown} follows Eq. (\ref{linear-g1-fit-eqn}): the left-hand side, calculated from the data, is fitted to a polynomial of a few terms in delay time $\tau$, hence providing estimates of $\beta$, $\bar{\Gamma}$, $\mu_2$ etc.  This method has the advantage that least-squares fitting to a polynomial which is linear in the unknown coefficients is a soluble problem that does not require iteration in the computer program \cite{bevington}.  A disadvantage of the method is that, to keep the higher-order terms in Eq. (\ref{linear-g1-fit-eqn}) small, the data have to be truncated at around $\bar{\Gamma} \tau = 1$(i.e. only data for $ \bar{\Gamma} \tau \leq 1 $ are kept) and it is not straightforward to determine the optimum truncation.  There is, in fact, a trade-off between large random errors in the fitted parameters if the data are truncated at too small a value of $\bar{\Gamma} \tau $ and large systematic errors (but smaller random ones) if too much of the data are used \cite{brown}.

Later, following rapid development of computer power, Frisken \cite{frisken} pointed out that iterative, non-linear fitting of 
\begin{equation}
\fl g^{(2)}(\tau) = B + \beta \left\{ \exp(-\bar{\Gamma} \tau) \left[
1
+ \frac{1}{2} \mu_2 \tau^2
- \frac{1}{3!} \mu_3 \tau^3
+ \frac{1}{4!} \mu_4 \tau^4
- \dots
\right] \right\}^2,
\label{fit-eqn}
\end{equation}
obtained from Eqs. (\ref{siegert-eqn}) and (\ref{g1-moments-eqn}), is a more robust procedure.  This approach has several advantages over the linear method.  First, it is not necessary to truncate the data since the divergence of the higher-order terms in Eq. (\ref{fit-eqn}) is suppressed by the decaying exponential pre-factor.  Second, the method allows the ``baseline'' $B$ to be regarded as a parameter to be fitted.  In an ideal experiment, $B$ should be 1 (Eq. (\ref{siegert-eqn})).  In practice $B$ can differ slightly from 1.  For example, slow drift of the laser intensity or of the gain of the detector leads to a spurious correlation, $B > 1$, which is almost independent of time over the span of the data.

Frisken \cite{frisken} used Eqs. (\ref{linear-g1-fit-eqn}) and (\ref{fit-eqn}) to analyze experimental data and clearly demonstrated the advantages of the non-linear method outlined above.  Subsequently Hassan and Kulshreshtha \cite{hassan} performed a similar analysis of experimental data and also considered simulated data for known distributions of decay rate $G(\Gamma)$.  However they only included terms up to second order in time and their simulated data did not take account of the uncertainty (noise) that is inevitable in an experiment.   In this paper we compare the two methods of analysis using simulated data with realistic noise over a range of polydispersities, with the aim of determining more precisely the potential and limitations of the non-linear method.

A note on terminology: Koppel \cite{koppel} pointed out that, formally, the logarithm of $g^{(1)}(\tau)$, Eq. (\ref{poly-g1-eqn}), is the cumulant generating function \cite{kendall} for the distribution $G(\Gamma)$.  Thus the expansion of this quantity, Eq. (\ref{linear-g1-fit-eqn}), is a power series in which the coefficients are the cumulants of $G(\Gamma)$; it is from this observation that the linear analysis based on Eq. (\ref{linear-g1-fit-eqn}) gets its commonly-used name, the ``method of cumulants''.  However, in the non-linear analysis using Eq. (\ref{fit-eqn}) emphasized here, it is the central moments rather than the cumulants that are relevant; thus what, to follow the custom, we have here called non-linear cumulant analysis might more logically be called the ``method of moments''.  (In fact, comparing Eqs. (\ref{linear-g1-fit-eqn}) and (\ref{fit-eqn}), we see that the cumulants only differ from the central moments at order 4 and higher \cite{kendall}.)  

\section{Methods}
\label{methods-sec}

\begin{figure}[htbp]
\begin{center}
\includegraphics[width=\columnwidth]{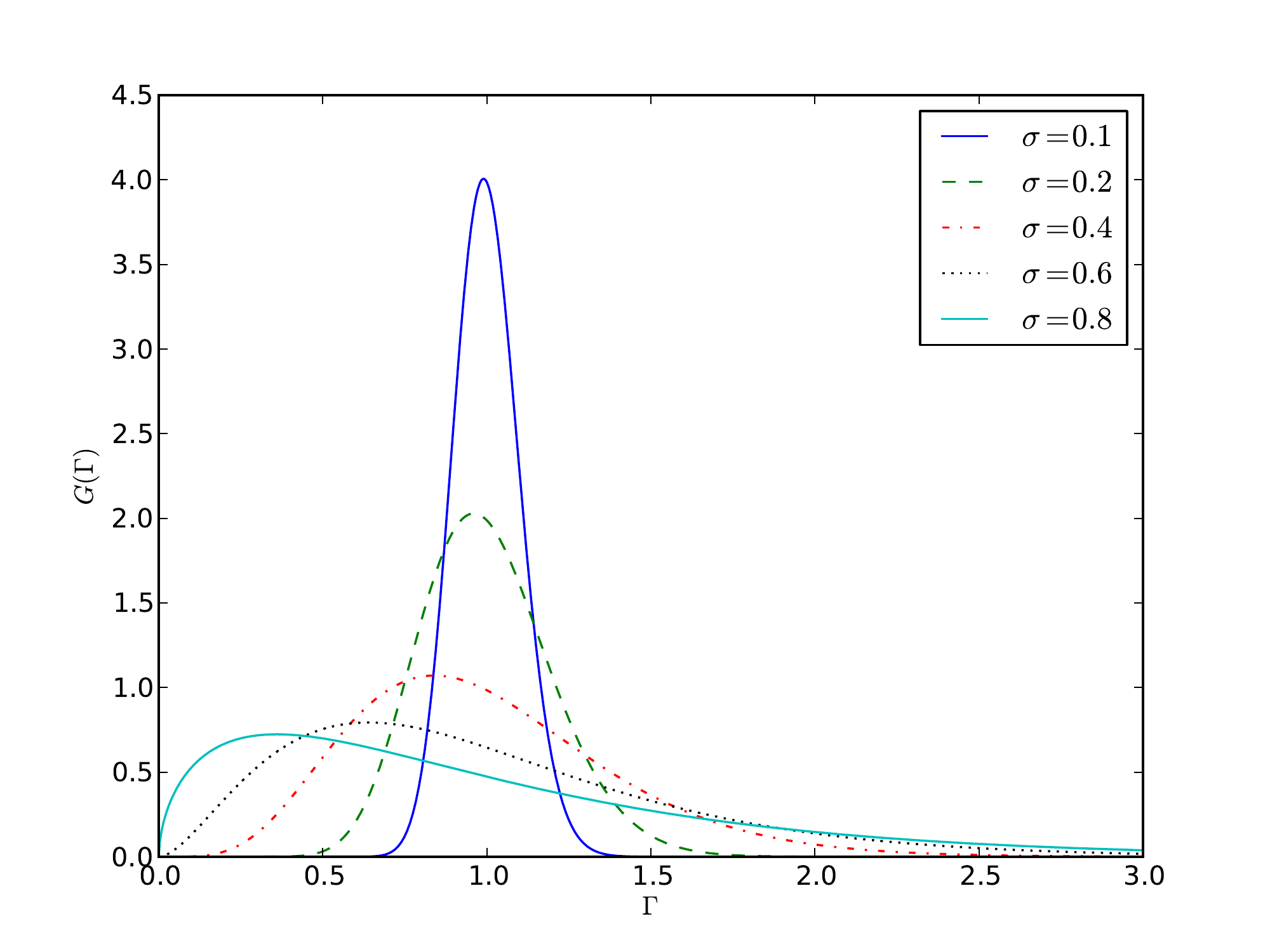}
\caption{The Schulz distribution, Eq. (\ref{schultz-eqn}), for indicated values of standard deviation $\sigma$.}
\label{schultz-fig}
\end{center}
\end{figure}

Mainly for mathematical convenience, we assume a Schulz distribution of decay rates:
\begin{equation}
G(\Gamma) = \frac{1}{\bar{\Gamma}} \frac{(z+1)^{z+1}}{z!} \left(\frac{\Gamma}{\bar{\Gamma}}\right)^{z} \exp \left( - \frac{\Gamma}{\bar{\Gamma}} (z + 1) \right);
\label{schultz-eqn}
\end{equation}
this is a two-parameter distribution defined by mean decay rate $\bar{\Gamma}$ and width $\sigma$ (normalized standard deviation), given by
\begin{equation}
\sigma^2 \equiv \frac{\overline{\Gamma^2} - \bar{\Gamma}^2}{\bar{\Gamma}^2} = \frac{1}{z+1}.
\end{equation}
Sample plots of the Schulz distribution are shown in Fig. (\ref{schultz-fig}).

Substitution of Eq. (\ref{schultz-eqn}) into Eq. (\ref{poly-g1-eqn}) gives
\begin{equation}
g^{(1)}(\tau) = \left(1 + \sigma^2 \bar{\Gamma}\tau\right)^{- 1 / \sigma^2};
\label{schultz-g1-eqn}
\end{equation}
(series expansion verifies that Eq. (\ref{schultz-g1-eqn}) reduces to a single exponential, Eq. (\ref{mono-g1-eqn}), as the width $\sigma$ of the distribution tends to zero).  The moments about the origin of the Schulz distribution are 
\begin{equation}
\fl \overline{\Gamma^n} \equiv \int{ \Gamma^n G(\Gamma d\Gamma) }
= \bar{\Gamma}^n ( 1 + (n - 1) \sigma^2 ) ( 1 + (n - 2) \sigma^2 ) \dots (1 + \sigma^2) ,
\end{equation}
giving, for the central moments, Eq. (\ref{mun-eqn}), 
\begin{equation}
\frac{\mu_2}{\bar{\Gamma}^2} = \sigma^2 \textrm{, } 
\frac{\mu_3}{\bar{\Gamma}^3} = 2 \sigma^4 \textrm{ and }
\frac{\mu_4}{\bar{\Gamma}^4} = 3 \left(\sigma^4 + 2 \sigma^6 \right) ;
\label{moments-eqn}
\end{equation}
the fourth cumulant of the distribution is 
\begin{equation}
\frac{\mu_4}{\bar{\Gamma}^4} - 3 \left( \frac{\mu_2}{\bar{\Gamma}^2} \right)^2 = 6 \sigma^6.
\end{equation}

Synthetic ``data'' $g^{(2)}(\tau)$ were constructed for a range of distribution widths $0 \leq \sigma \leq 1$ from
\begin{equation}
g^{(2)}(\tau) = B + \beta \left[ g^{(1)}(\tau) \right]^2
\label{sim-g2-eqn}
\end{equation}
with $g^{(1)}(\tau)$ given by Eq. (\ref{schultz-g1-eqn}); here $B$ is the baseline, cf. Eq. (\ref{fit-eqn}).   To mimic modern photon correlators, 150 data points were logarithmically spaced in scaled delay time $\bar{\Gamma} \tau$ in the range $10^{-2} \leq \bar{\Gamma} \tau \leq 10^2$.  To mimic experimental noise, each value of $g^{(2)}(\tau)$ was multiplied by a random number drawn from a Gaussian distribution of mean 1 and standard deviation $s$.  For most of the analysis we took $s = 10^{-3}$, corresponding to an uncertainty of one part in a thousand on each data point.  This is the typical magnitude of counting errors in a DLS experiment.  We also looked briefly at noisier data, up to $ s = 10^{-2} $.  For each value of $\sigma$, 20 data sets with different random noise were analyzed, allowing calculation of the means and standard deviations of the fitted parameters.

Two analyses of the data were performed.  In the standard cumulant analysis, we assumed $B$ to take its ideal value of 1 in Eq. (\ref{sim-g2-eqn}).  Then $\ln \sqrt{g^{(2)}(\tau) - 1 }$, calculated from Eq. (\ref{siegert-eqn}), was fitted by linear least squares to Eq. (\ref{linear-g1-fit-eqn}) \cite{bevington}.  Fits to polynomials in $\tau$ of order one, two, three and four were performed, providing estimates of an increasing number of the moments $\mu_n$.  For simplicity, both $\bar{\Gamma}$ and $\beta$ were assumed to be 1 when generating the data, but were taken as parameters to be fitted in the analysis.  In this linear cumulant analysis, the data were truncated when $g^{(2)}(\tau) - 1$ dropped to 10\% of its initial value.

In the second, non-linear, analysis, the simulated data for $g^{(2)}(\tau)$ were fitted to Eq. (\ref{fit-eqn}) using a variable metric method \cite{james}.  As with the standard analysis, four orders of fit were performed.  The data were not truncated and the background $B$ was regarded as an additional floating parameter.

\section{Results}
\label{results-sec}

\begin{figure}[htbp]
\begin{center}
\includegraphics[width=0.75\textwidth]{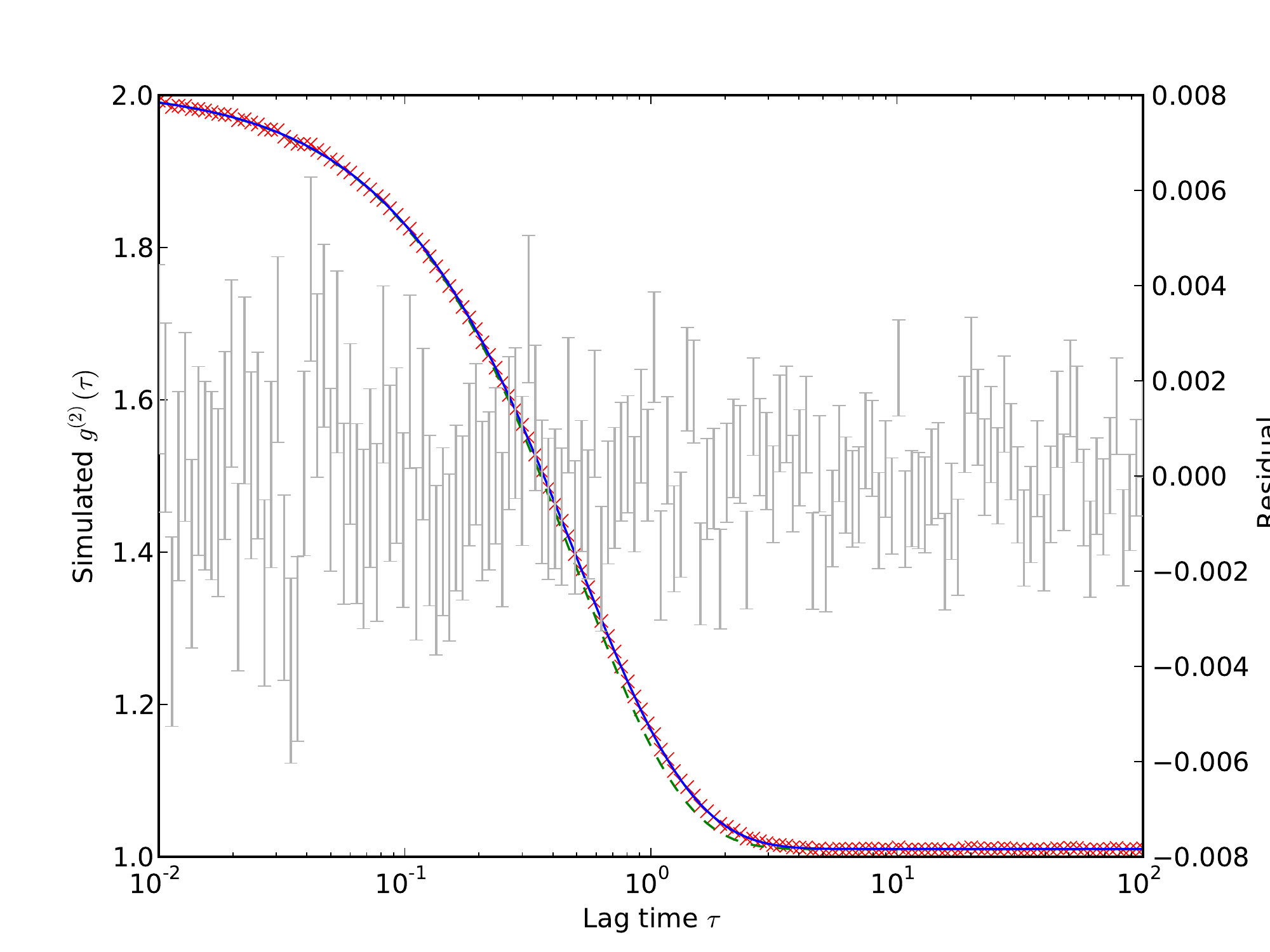} \includegraphics[width=0.75\textwidth]{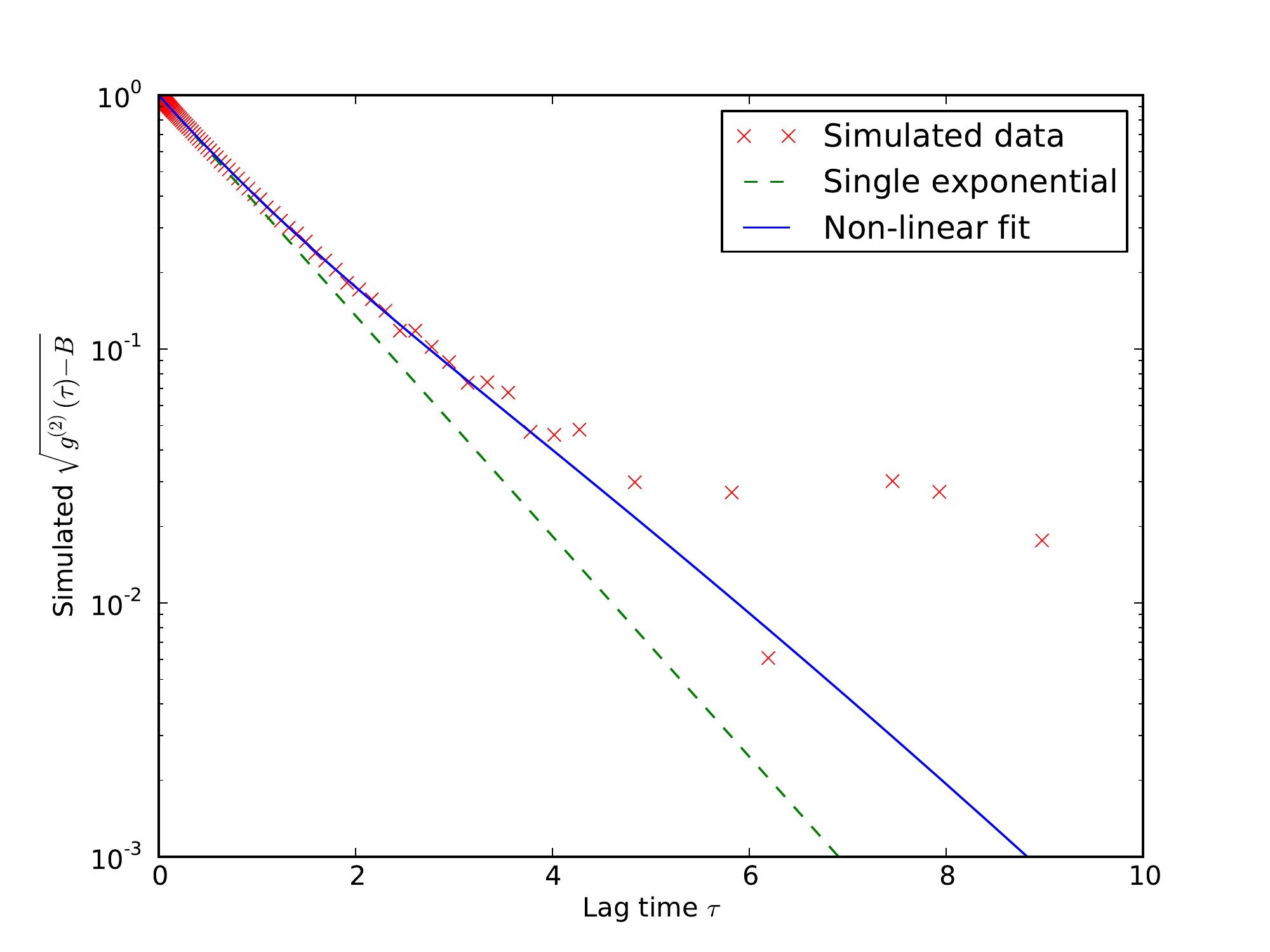}
\caption{Example of a fit to simulated data.  \emph{(a, top)} Crosses: simulated data, Eqs. (\ref{sim-g2-eqn}) and (\ref{schultz-g1-eqn}), with $B = 1.01, \beta = 1, \bar{\Gamma} = 1 , \sigma = 0.4$ and noise $s = 10^{-3}$.  Solid line: fourth-order non-linear fit of the data (Eq. (\ref{fit-eqn})).  Residuals, difference between data and fit, are indicated.  Dashed line; single-exponential decay with $\bar{\Gamma} = 1$.  \emph{(b, bottom)} Semi-logarithmic representation of the same data.}
\label{sim-res-fig}
\end{center}
\end{figure}

Figure \ref{sim-res-fig}(a) shows an example of data fitted successfully by the fourth-order non-linear procedure.  Input values were $\sigma = 0.4$ and $B = 1.01$, and all the fitted parameters are within the expected uncertainty of the input values.  The dashed line shows a single exponential with decay rate $\bar{\Gamma} = 1$.  The figure illustrates how even a significant spread of particle size (see Fig. \ref{schultz-fig}), $\sigma = 0.4$, leads to a correlation function that does not differ much from a single exponential.  In the semi-logarithmic representation of Fig. \ref{sim-res-fig}(b), the difference, at larger delay times, is more apparent.

\subsection{Mean decay rate}

\begin{figure}[htbp]
\begin{center}
\includegraphics[width=\textwidth]{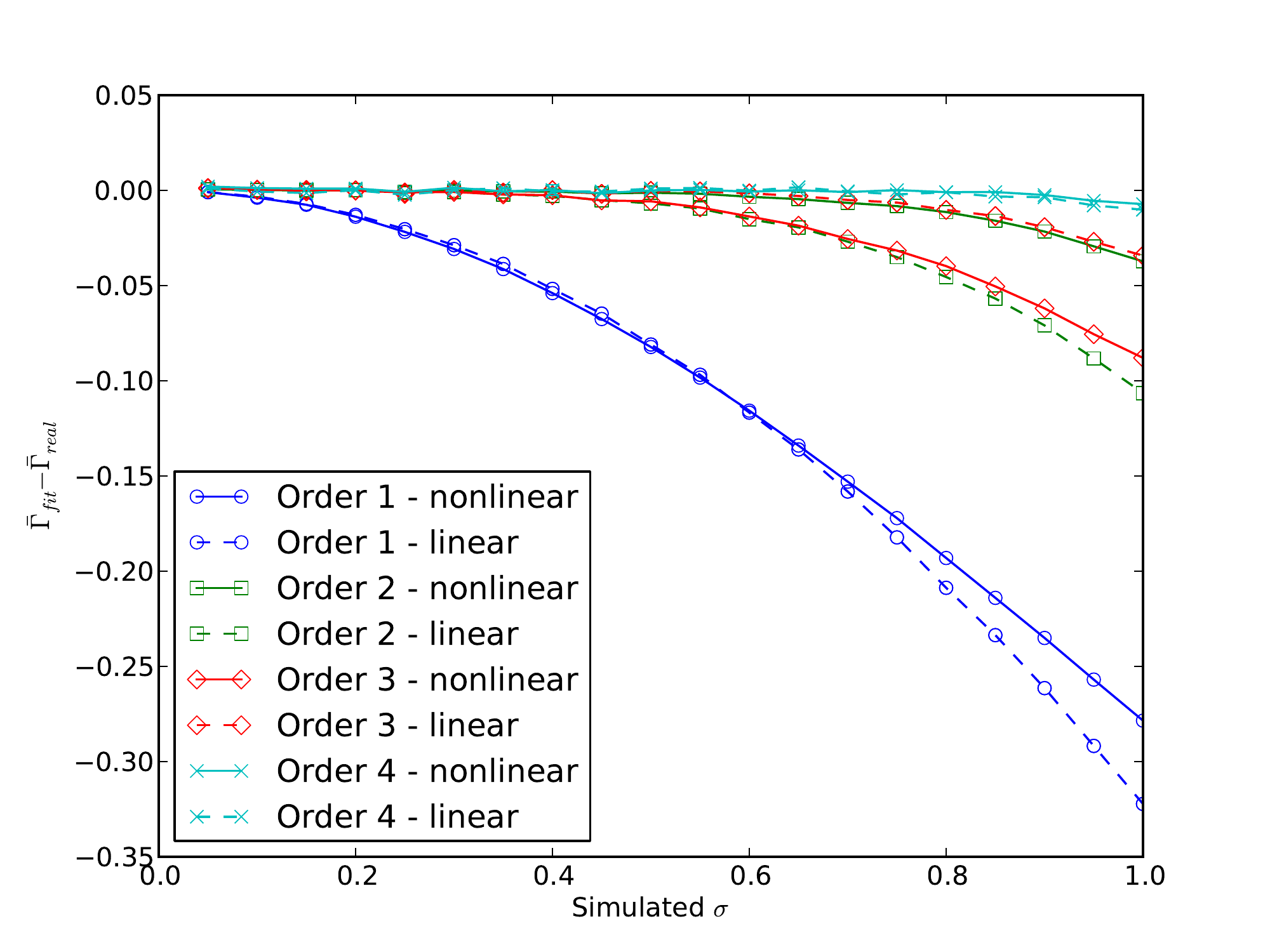}
\caption{Deviation of the fitted mean decay rate $\bar{\Gamma}$ from its input value, 1, as a function of polydispersity $\sigma$ for both linear and non-linear first- to fourth-order fits.  The error bar indicates the standard deviation of the fitted $\bar{\Gamma}$ for the fourth-order non-linear fit across the 20 generated data sets at $\sigma=0.4$.  See text for discussion.}
\label{gamma-dev-fig}
\end{center}
\end{figure}

Figure \ref{gamma-dev-fig} shows the deviation of the fitted mean decay rate $\bar{\Gamma}$ from its input value, 1, as a function of polydispersity $\sigma$ for both linear and non-linear first- to fourth-order fits.  A fit of order 1 is the equivalent of force-fitting the data to a single exponential.  It is clear that, as soon as polydispersity becomes significant, the first-order fits seriously underestimate $\bar{\Gamma}$.  However, adding just one parameter, $\mu_2$, in the second-order fits immediately allows reliable estimates of $\bar{\Gamma}$ up to polydispersities of 0.4 to 0.5; fourth-order fits estimate $\bar{\Gamma}$ reliably over almost the whole range of polydispersity considered.  The error bar in Fig. \ref{gamma-dev-fig} indicates the standard deviation of the fitted $\bar{\Gamma}$ for the fourth-order non-linear fit across the 20 generated data sets at $\sigma = 0.4$, and shows that $\bar{\Gamma}$ can be obtained with an accuracy of better than 0.5\% for moderately polydisperse samples.

In the estimation of $\bar{\Gamma}$ for moderately polydisperse systems, there is little to choose between non-linear and (truncated) linear fits.  Intriguingly, at third-order, the non-linear fit actually does worse than the linear one.  We note that, for a symmetrical distribution of decay rates, the odd-order central moments are zero.  Thus, in general, $\mu_4 \tau^4$ can be larger than $\mu_3 \tau^3$ in Eq. (\ref{fit-eqn}) even at $\bar{\Gamma} \tau < 1$, and there is no justification for a fit that includes  $\mu_3$ but not $\mu_4$.

\subsection{Second moment}

\begin{figure}[htbp]
\begin{center}
\includegraphics[width=\textwidth]{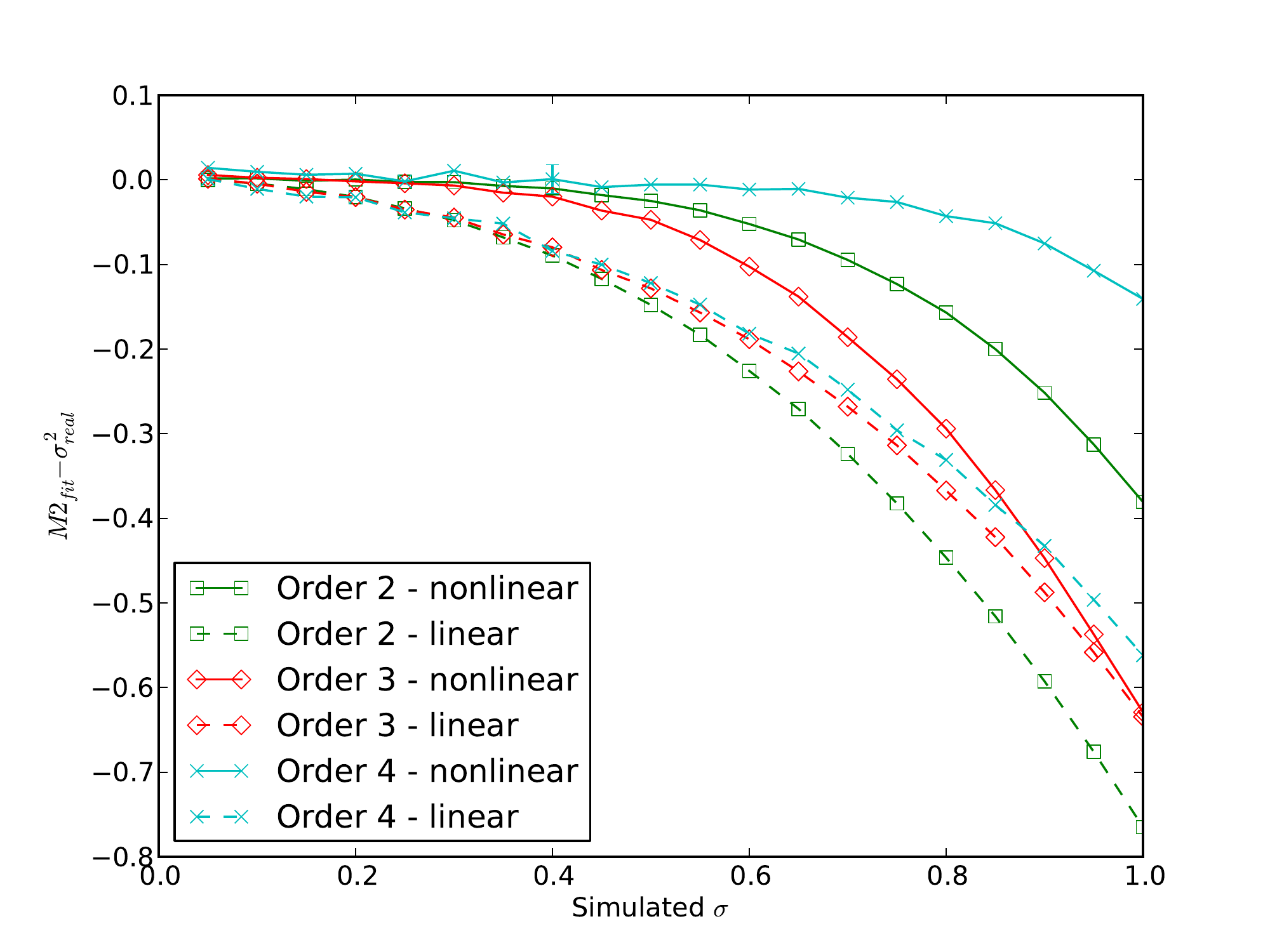}
\caption{Deviation of the fitted second central moment $\mu_2$ from its input value $\sigma^2$ (Eq. (\ref{moments-eqn}) with $\bar{\Gamma} = 1$) for linear and non-linear fits up to fourth order.  Similar to Fig. \ref{gamma-dev-fig}, the error bar indicates the standard deviation of the fitted $\mu_2$ for the fourth-order non-linear fit across the 20 generated data sets at $\sigma=0.4$.}
\label{mu2-dev-fig}
\end{center}
\end{figure}

Under the same conditions as for Fig. \ref{gamma-dev-fig}, Fig. \ref{mu2-dev-fig} shows the deviation of the fitted second central moment $\mu_2$ from its input value $\sigma^2$ (Eq. (\ref{moments-eqn}) with $\bar{\Gamma} = 1$).  For this parameter, all three non-linear fits do much better than the linear ones, giving accurate estimates of $\mu_2$  for polydispersities up to 0.3.  The fourth-order non-linear fit gives a good estimate of $\mu_2$ up to about $\sigma = 0.6$.  At $\sigma = 0.4$, where $\mu_2 = \sigma^2 = 0.16$, uncertainty in the determination of $\mu_2$ is about $\pm 0.02$, a finding consistent with experience in experiments \cite{brown}.  These values translate into a quite accurate determination of the width, $\sigma(\textrm{fitted}) = 0.40 \pm 0.025$

\subsection{Third moment}

\begin{figure}[htbp]
\begin{center}
\includegraphics[width=\textwidth]{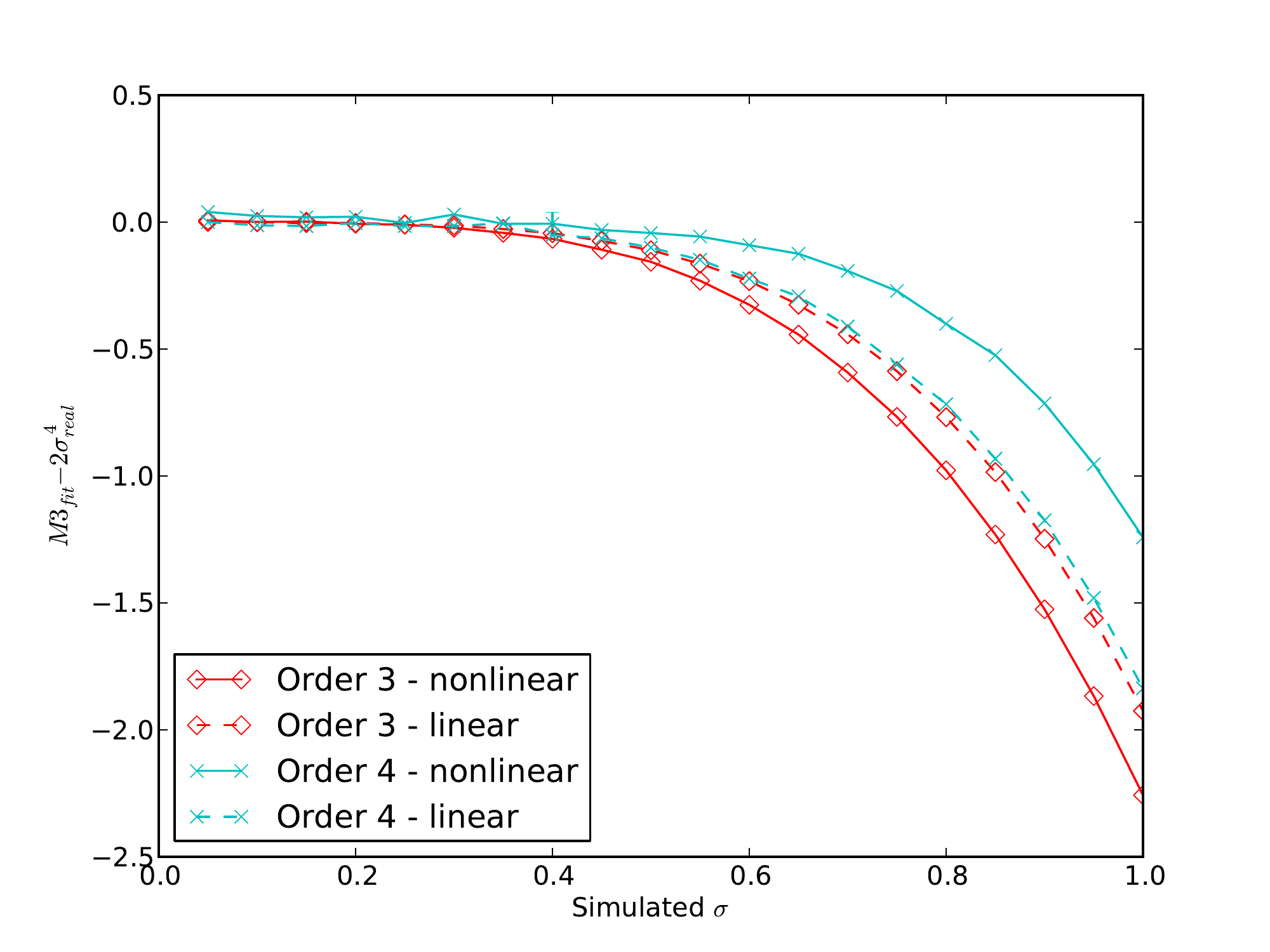}
\caption{Deviation of the fitted third central moment $\mu_3$ from its input value $2\sigma^4$ (Eq. (\ref{moments-eqn}) with $\bar{\Gamma} = 1$).}
\label{mu3-dev-fig}
\end{center}
\end{figure}

Figure \ref{mu3-dev-fig} shows the deviation of the third moment from its input value $2\sigma^4$ (Eq. (\ref{moments-eqn}) with $\bar{\Gamma} = 1$).  We have mentioned above that there is no justification for a third-order fit.  At first sight the fourth-order fit appears to do well up to $\sigma = 0.4$.  However we note that, for the fourth-order fit at $\sigma = 0.4$, the uncertainty, $\pm 0.045$, in the recovered $\mu_3$ is barely smaller than the value, 0.0512, of $\mu_3$ itself.  Thus, at least for this distribution and noise level, DLS can do little more than hint at the sign of $\mu_3$.

\subsection{Fourth moment}

We find that, up to about $\sigma = 0.4$, the uncertainty in the fitted value of $\mu_4$ is comparable to its actual value.  On the other hand, above $\sigma = 0.4$, as $\mu_4$ becomes larger, its fitted value is consistently lower than its actual value.  Thus, as with $\mu_3$, we obtain little useful information.

\subsection{Baseline}

Provided that the data extend to long enough times, as in Fig. \ref{sim-res-fig}, we find that the baseline $B$ is recovered accurately by the non-linear fits.  Furthermore, treating the baseline as a parameter to be fitted introduces almost no further uncertainty in the other fitted parameters (compared to the case where the baseline is fixed at its input value). 

\subsection{Initial guesses}

The non-linear fitting procedure, Eq. (\ref{fit-eqn}), appears to be very stable with respect to the choice of initial values for the fitted parameters.  The simplest approach is to obtain initial values of $\beta$ and $\bar{\Gamma}$ from a first-order linear fit of a short-time portion of the data, to assume that the initial value of the baseline $B$ is 1, and to take initial values for the higher moments, $\mu_2, \mu_3,$ etc., to be zero.

\subsection{Larger noise}

With noisier data, $s = 10^{-2}$ (see section 3), fourth-order fits were not useful, giving large uncertainties in both $\bar{\Gamma}$ (several percent) and $\mu_2$ (typically $\pm 0.20$).  However second-order non-linear fits gave $\bar{\Gamma}$ to within about 2\% with an uncertainty in $\mu_2$ of about $\pm 0.05$, meaning that polydispersities $\sigma \left( = \sqrt{\mu_2} \right)$ greater than 0.20 to 0.25 could still be detected.

\section{Discussion}
\label{discuss-sec}

Several conclusions may be drawn from these results.  First, we find that non-linear fits are more accurate and more straightforward to perform than linear fits (where a somewhat arbitrary truncation of the data is necessary).  This is seen clearly in Fig. \ref{mu2-dev-fig} which shows that non-linear fits return a much smaller systematic error in the second moment $\mu_2$ than linear fits.  Second, we have pointed out that there is no justification for performing third-order fits, either linear or non-linear: the $\tau^4$-term in Eqs. (\ref{linear-g1-fit-eqn}) and (\ref{fit-eqn}) can, in general, be comparable to or larger than the $\tau^3$-term even at small $\bar{\Gamma}\tau$.  Third, performing a fourth-order, rather than a second-order, fit reduces the systematic error in the fitted $\mu_2$ (Fig. \ref{mu2-dev-fig}).  However, even for polydispersities as large as $\sigma = 0,6$, fourth-order fits do not provide reliable estimates of the higher moments, $\mu_3$ and $\mu_4$.  This finding agrees with experience: we are not aware of any experiment where $\mu_3$ or $\mu_4$ have been convincingly measured.  

It might at first appear surprising that including unknown parameters $\mu_3$ and $\mu_4$ in the non-linear fit should improve the fitting of $\mu_2$ even though the determined values of $\mu_3$ and $\mu_4$ are themselves unreliable. Comparisons of the $\chi^2$ surface \cite{bevington} in response to combined changes of $\mu_2$ with $\bar{\Gamma}$ and $\mu_2$ with $\mu_3$ reveal that the additional fit parameters significantly reduce the correlations between $\mu_2$ and $\bar{\Gamma}$ i.e. the minimum area in the $\chi^2$ surface becomes perpendicular to either the $\mu_2$ or the $\bar{\Gamma}$ axes when $\mu_3$ and $\mu_4$ are included in the fit. Without these parameters the $\chi^2$ minimum is typically aligned parallel to $\mu_2 = \bar{\Gamma}$ making it more difficult to determine the optimal fit values.

Despite the somewhat negative conclusions of the first paragraph of this section, we \emph{are} able to suggest a robust and relatively straightforward procedure to obtain the first two moments of the distribution of diffusion constants, $\bar{\Gamma}$ and $\mu_2$:

\begin{enumerate}

\item Set up a photon correlator with channels logarithmically-spaced in time so that the measured intensity correlation function $g^{(2)}(\tau)$ extends well into the long-time baseline $B$; run the experiment for long enough that the typical noise on a data point is about 1 part in $10^3$.

\item Obtain initial estimates of the coherence factor $\beta$ and mean decay rate $\bar{\Gamma}$ from a short-time linear fit of the data, and take other initial estimates as baseline $B = 1$, higher moments $\mu_2, \mu_3, \mu_4 = 0$.

\item Perform a \emph{fourth}-order non-linear fit of  $g^{(2)}(\tau)$ to Eq. (\ref{fit-eqn}). If the fitted second moment $\mu_2 / \bar{\Gamma}^2$ is less than about 0.40 (corresponding to $\sigma \approx 0.6$ in Fig. 4), then the fitted $\bar{\Gamma}$ and $\mu_2 / \bar{\Gamma}^2$ should be good estimates of the true values, accurate to around 0.5\% and $\pm 0.02$ respectively.

\item If, in (iii), the fitted second moment $\mu_2 / \bar{\Gamma}^2$ is less than $\sim 0.1$ (polydispersity $\sigma \leq 0.3$), then more precise estimates of the values of $\bar{\Gamma}$ and $\mu_2 / \bar{\Gamma}^2$ should be obtainable from a \emph{second}-order non-linear fit of $g^{(2)}(\tau)$ to Eq. (\ref{fit-eqn}) (see Figs. 3 and 4).

\end{enumerate}

Here we have only considered a Schulz distribution of decay rates $G(\Gamma)$ (Eq. (\ref{schultz-eqn})). However it is straightforward to show that the moments of a distribution can be written~\cite{Krieger, Pusey-Fijnaut-Vrij}
\begin{equation}
\overline{\Gamma^n} = \overline{\Gamma}^n \left[1 + \frac{n(n-1)}{2}\sigma^2 + O(\sigma^3)\right].
\label{narrow-eqn}				
\end{equation}

\noindent For, fairly narrow, fairly symmetrical distributions it is sufficient to keep only the first two terms so that the higher moments can, to a good approximation, be written purely in terms of the standard deviation $\sigma$.  Thus we may expect that the general discussion above will apply to any distribution which is narrow enough.  For the Schulz distribution, we have shown here that ``narrow enough'' means $\sigma$ less than about 0.6; but, for more skewed or flatter (larger kurtosis) distributions than the Schulz, this upper limit might need to be reduced somewhat.  Repeating the programme of this paper for other distributions such as the lognormal or a two-component mixture would provide more quantitative conclusions, but the general picture is unlikely to change.

The method of cumulants described here is relatively simple to implement and does not require the input of any prior information about the sample.  It should therefore be the first approach of any experimenter presented with an unknown sample.  When this first measurement returns a large enough second moment, it is probably worth trying some of the more complex analysis methods listed in section 1.  Experience with these methods~\cite{Finsy} shows that they have difficulty resolving bimodal distributions when the ratio of the two sizes is less than about 3.  This corresponds to a standard deviation $\sigma \approx 0.5$ and a second moment $\mu_2 / \bar{\Gamma}^2 \approx 0.25$. In general, though, these methods work best for significantly broader distributions than this.

In this paper we have limited consideration to analysis of a single DLS measurement.  For completeness, we mention that, for particles large enough that there is significant angular variation in the intensity of light that they scatter (radius greater than about 50 nm), performing a combined analysis of data taken at different scattering angles can frequently provide much more detailed information.  For example, by measuring the angular dependence of the apparent average diffusion constant of particles which show a minimum in their angular intensity profile, it is possible to measure very small polydispersities~\cite{Pusey&VanMegen}.  It has recently been demonstrated that a Bayesian analysis of the full correlation functions measured at several scattering angles can be very powerful in resolving multi-modal distributions~\cite{Naiim}.

We note that obtaining information about the distribution $G(\Gamma)$ of decay rates (or diffusion constants) is usually not the ultimate goal of an analysis of DLS data; rather, one is interested in the distribution of particle sizes.  Because the contribution of each particle species to $G(\Gamma)$ is weighted by the intensity of light scattered by that species and because the particle radius is, through Eqs. (\ref{diff-eqn}) and (\ref{last-dls-eqn}), inversely proportional to the decay rate $\Gamma$, obtaining the size distribution from $G(\Gamma)$ is not straightforward.  Nevertheless, for many systems, such as solid spheres \cite{pusey1}, spherical shell-like particles \cite{patty} and random-coil polymers \cite{brown}, the measured moments $\bar{\Gamma}$ and $\mu_2 / \bar{\Gamma}^2$ can be related to moments of the particle size distribution.  For example, for homogeneous spheres much smaller than the wavelength of light, it can be shown \cite{pusey1, Pusey&VanMegen} that the effective particle radius obtained by substituting the measured $\bar{\Gamma}$ in Eqs. (\ref{diff-eqn}) and (\ref{last-dls-eqn}) is
\begin{equation}
R_{\textrm{eff}} \left[ = \frac{k_B T q^2}{6 \pi \eta \bar{\Gamma}} \right] = \overline{R^6} / \overline{R^5},
\end{equation}
and that 
\begin{equation}
\frac{\mu_2}{\bar{\Gamma}^2} = \frac{\overline{R^4} \, \overline{R^6}}{(\overline{R^5})^2} - 1,
\end{equation}
where the $\overline{R^n}$ are moments of the distribution of particle radii.  For narrow size distributions, where an approximation like that of Eq. (\ref{narrow-eqn}) can be applied, these results reduce to the simpler (and useful) expressions $R_{\textrm{eff}} = \overline{R} \left( 1 + 5\sigma^2_R\right)$ and $\mu_2 / \bar{\Gamma}^2 = \sigma_R^2$ where $\sigma_R$ is the standard deviation of the radius.

Finally we comment that the conclusions reached here --- useful determination of $\bar{\Gamma}$ and $\mu_2$, but little prospect of obtaining higher moments --- are not too different from those reached by Koppel \cite{koppel} when introducing the linear cumulant method more than 40 years ago.  However we have shown that non-linear fitting gives a simpler and more robust procedure which avoids the necessity of a rather arbitrary truncation of the data; we have estimated the upper limit of reliability of cumulants methods at about $\mu_2 / \bar{\Gamma}^2 = 0.4$, polydispersity $\sigma \approx 0.6$; and we have suggested that if an initial analysis by the cumulants method yields a second moment more than about $\mu_2 / \bar{\Gamma}^2 = 0.25$, it is probably worth trying to obtain more information from one of the more complex analysis methods, based on inverse Laplace transformation, mentioned above.

\section*{References}
\bibliography{refs}
\bibliographystyle{unsrt}

\end{document}